\documentclass[doublecol]{epl2}
\usepackage{amsmath}
\usepackage{amssymb}
\usepackage{graphicx} 
\usepackage{hyperref}
\usepackage{color}

\newcommand{\opgamma}[1]{{\hat{\gamma}^{\phantom \dagger}}_{#1}}
\newcommand{\opgammadag}[1]{{\hat{\gamma}^{\dagger}}_{#1}}

\newcommand{\ket}[1]{\left|#1\right\rangle}
\newcommand{\bra}[1]{\left\langle #1\right|}
\newcommand{\ud}{\mathrm{d}}

\newcommand{\nep}{\textrm{e}}

\newcommand{\sign}{\operatorname{sign}}
\newcommand{\opcdag}[1]{{\hat{c}^{\dagger}}_{#1}}
\newcommand{\opc}[1]{{\hat{c}^{\phantom \dagger}}_{#1}}

\begin{document}

\title{Kibble-Zurek Scaling in Periodically-Driven Quantum Systems}

\author{Angelo Russomanno\inst{1,2,3}\and Emanuele G. Dalla Torre\inst{1}}
\shortauthor{A. Russomanno and E.~G. Dalla Torre}

\institute{
\inst{1} Department of Physics, Bar Ilan University, 52900, Ramat Gan Israel \\
\inst{2} Scuola Normale Superiore, Piazza dei Cavalieri 7, I-56127 Pisa, Italy \\
\inst{3} International Centre for Theoretical Physics (ICTP), Strada Costiera 11, I-34151 Trieste, Italy
}

\abstract{
We study the slow crossing of non-equilibrium quantum phase transitions in periodically-driven systems. We explicitly consider a spin chain with a uniform time-dependent magnetic field and focus on the Floquet state that is adiabatically connected to the ground state of the static model. We find that this {\it Floquet ground state} undergoes a series of quantum phase transitions characterized by a non-trivial topology. To dynamically probe these transitions, we propose to start with a large driving frequency and slowly decrease it as a function of time. Combining analytical and numerical methods, we uncover a Kibble-Zurek scaling that persists in the presence of moderate interactions. This scaling can be used to experimentally demonstrate non-equilibrium transitions that cannot be otherwise observed.
}


\maketitle

%
Recent 
experiments on
artificial quantum many-body systems 
demonstrated quantum-coherent evolution under various
external perturbations~\cite{Houck_Nat,Bloch_RMP08,Bloch_Nat,Houck_Nat,Polkovnikov_RMP11} and 
triggered the study of fundamental aspects of quantum non-equilibrium physics. 
One important direction is the quest for
universal properties in the 
dynamics~\cite{Polkovnikov_RMP11,Emanuele_PRL13,russomanno_JSTAT15,Zurek_PRL05,Grandi_PRB10,Berges_PRL15,Vasseur_PRX14,Latta_Nat15}. 
A universal non-equilibrium behaviour can originate from the universality of an underlying equilibrium problem~\cite{Emanuele_PRL13,russomanno_JSTAT15}: A well known example is offered by
the Kibble-Zurek (K-Z) scaling observed when crossing a quantum phase transition
~\cite{Polkovnikov_PRB2005,Zurek_PRL05,Polkovnikov_RMP11,Sachdev:book,Campo_PRA14}. In addition, recent studies found evidence of genuine non-equilibrium universality following a quantum quench~\cite{Grandi_PRB10,Berges_PRL15,Vasseur_PRX14,Latta_Nat15}, or in presence of a time-dependent noise~\cite{Emanuele_Nat10,jamir_15}.
%
Another important direction is the exploration of the conditions for an asymptotic steady state~\cite{Greiner_Nat02bis,Kinoshita_Nat06,Polkovnikov_RMP11,Barthel:PRL08,Rigol_Nat,Rigol_PRL07,Kollath_PRL,Manmana_PRL,Canovi_PRB11,Marquardt_PRE12,Kollar_PRA08,Eckstein_PRL08,Cazal:PRL06,Cazal:PRA09,Calabrese_JSTAT07,Calabrese_JSTAT12a,Calabrese_JSTAT12b,Cramer_PRL08,
Cramer_PRL08b,Trotzky:NP12,Cramer_NJP12,Ziraldo_PRL12,art:Ziraldo_PRB13,Santos_PRE10,Sknappa_PRB15}.
In the context of quantum quenches, it was found that systems with a continuous spectrum generically relax to a diagonal ensemble~\cite{Ziraldo_PRL12,
art:Ziraldo_PRB13}. 
 {Current investigations} focus
on understanding when this asymptotic condition is thermal (ergodic dynamics) and
when it is not (regular dynamics)~\cite{Marquardt_PRE12,Canovi_PRB11,Santos_PRE10,Manmana_PRL,Rigol_Nat,Rigol_PRL07}.
The transition from regular to ergodic dynamics is
connected to the emergence of quantum chaos~\cite{alessio_15,Haake:book,Casati_PTPS89} and to the
many-body localization/delocalization transition~\cite{Alt_PRL97,Basko_Ann06,Pal_Huse_PRB10,Ehud_preprint:15,Marquardt_PRE12,Canovi_PRB11,Santos_PRE10,Casati_PRL90}.
These studies are 
deeply related to the main problem of statistical mechanics:
how microscopic quantum coherent dynamics can induce macroscopic thermalization~\cite{Deutsch_PRA91,Sred_PRE94,Rigol_Nat,Santos_PRE10}.

%
In this work we address the universal scaling and the absence of thermalization in periodically-driven many-body systems~\cite{Lignier_PRL07,Zenesini_PRL09,Gemelke_PRL05,Eckardt_PRL051,Kollath_PRL06,Ponte_PRL15,Ponte_AP15,
Kim_PRE14,Polkovnikov_NatPhys11,Emanuele_2014:preprint,Dalessio_AP13,Rigol_PRX14,Russomanno_PRL12,Russomanno_JSTAT13,Russomanno_EPL15,
Lazarides_PRL14,Lazarides_PRE14,Batisdas_PRA12,mori_15,Knappa_preprint15,Lindner_Nat11,LS_15,Jotzu_Nat14,Oca_PRB79,Iacopo_preprint15,Buco_arXiv15,sciarmata_EPL14,russomanno_16}.
We focus on Floquet states, the eigenstates of the periodically driven dynamics. We specifically address a {Floquet state which shows
many properties analogous to the ground state in the static system: we term ``Floquet ground state'' (FGS). We emphasize that -- opposite to the GS of a static Hamiltonian -- the
FGS depends on time in a periodic fashion (up to a phase factor) and, in general, cannot be found as the minimum-energy eigenstate of any
Hamiltonian. Nevertheless it shows many properties similar to the GS of a static Hamiltonian. First of all, } 
it shows
an area-law entanglement entropy, in contrast to thermal steady states whose entanglement grows as a volume law~\cite{Ponte_PRL15}. Furthermore, 
the FGS can undergo second-order quantum phase transitions. 
In the case of an integrable spin-chain model, we show that these transitions are associated with the closing of a Floquet gap and separate non-equilibrium phases with distinct topology. Next, we discuss how to dynamically probe such transitions: in contrast to the static case, 
the Floquet ground state in general cannot be found either numerically or experimentally at an extremum of the spectrum of some Hamiltonian. We propose to reach the FGS by initially preparing the system at high frequency in the ground state of the time-averaged Hamiltonian -- which in the high frequency limit coincides with the FGS -- and then slowly decreasing the driving frequency. In this way we can adiabatically follow the FGS, until a non-equilibrium transition is crossed. As we discuss in detail, this is possible {thanks to the Floquet adiabatic theorem ~\cite{Messiah:book,Breuer_ZPD89,Young_70} and the fact that the FGS is the adiabatic continuation at finite frequency of the GS of the time-averaged Hamiltonian taken in the high frequency limit}. We discover that the crossing of the non-equilibrium transition is marked by a Kibble-Zurek scaling of physical
observables (we define this phenomenon ``Floquet Kibble-Zurek''), which can be used to determine the position and critical exponents of the transition. 

Because the entanglement entropy remains small during the dynamical crossing of the transition, this process can be efficiently studied with the time-dependent DMRG algorithm. We find that the Floquet-Kibble-Zurek scaling persists even in the presence of moderate integrability-breaking terms. It is important to stress that, in spite of the slowness of  the crossing, the {Floquet} adiabatic theorem 
does not directly apply. Due to the folding of the Floquet spectrum (described in detail later), interactions can in principle give rise to a large number of
exponentially small gaps which will prevent the observation of the transition. The persistence of the non-equilibrium transition is probably connected with its topological nature, which prevents these gaps from opening. We conjecture that in the presence of larger interactions a strong mixing of the Floquet states is expected, leading to Floquet states obeying Eigenstate Thermalization~\cite{Deutsch_PRA91,Sred_PRE94,Rigol_Nat,Lazarides_PRE14,Rigol_PRX14,Russomanno_EPL15}.
%
%
%
%

%
Floquet theory~\cite{Hanggi_book,Shirley_PR65, Hausinger_PRA10} is the basis of our analysis. 
For any time-periodic Hamiltonian, $\hat{H}\left(t\right)=\hat{H}\left(t+2\pi/\Omega\right)$ {(here and in the following, $\Omega$ is the frequency and
$\tau=2\pi/\Omega$ is the period)}, it is possible to find a complete basis of solutions of the Schr\"odinger equation 
(the Floquet states) which are periodic up to a phase $\ket{\Psi_{\gamma,\Omega}(t)}=\nep^{-i\bar{\mu}_\gamma t}\ket{\Phi_{\gamma,\Omega}(t)}$.
The many-body quasi-energies $\bar{\mu}_\gamma$ are real, and the Floquet modes $\ket{\Phi_{\gamma,\Omega}(t)}$ are periodic.
These states are analogous to the Bloch wave eigenfunctions of space-periodic 
Hamiltonians~\cite{Hausinger_PRA10,Grosso:book}.
Like the Bloch quasi-momenta, the quasi-energies are not unique and are defined up to translations of an integer number of $\Omega$~\cite{Shirley_PR65}.
In this way, the whole Floquet spectrum can be folded in any interval of length $\Omega$ which defines thereby a Brillouin zone.
Like eigenenergies, the quasi-energies can become degenerate: two of them taken in the same Brillouin zone are degenerate (or in resonance)
if they coincide or if they differ by $\Omega$. 
Each degeneracy in the Floquet spectrum corresponds to a many-photon resonance in the system~\cite{Shirley_PR65}. 
{Floquet modes at time\footnote{We can restrict to this time interval thanks to the $\tau$-periodicity of the Floquet modes.} $0\leq t\leq\tau$ and quasi-energies can be extracted diagonalizing the evolution operator over one period
$\hat{U}(t+\tau,t)$: we have 
$\hat{U}(t+\tau,t)=\exp\left(-i\hat{H}_{F,\Omega}(t)\tau\right)$} with the so-called ``Floquet Hamiltonian'' defined as 
$\hat{H}_{F,\Omega}(t)\equiv\sum_{\gamma}\overline{\mu}_\gamma\ket{\Phi_{\gamma,\Omega}(t)}\bra{\Phi_{\gamma,\Omega}(t)}$, where
$\overline{\mu}_\gamma$ are conventionally taken in the first Brillouin zone $[-\Omega/2,\Omega/2]$. 

%

{In this work the Floquet adiabatic theorem~\cite{Messiah:book,Breuer_ZPD89,Young_70} plays a very important role. 
We prepare the driven system at time 0 in a Floquet state $\ket{\Psi_{\gamma}(0)}$ and
we very slowly change (on a time scale $t_f$) some parameters ${\bf R}$ of the periodic driving (for instance, the frequency or the amplitude),
starting from a value ${\bf R}(0)$. 
Floquet adiabatic theorem states that, if the parameters are changed in time slowly enough, the dynamics transports the system 
after a time $t$ to a state $\ket{\Psi_{\gamma,\,{\bf R}(t)}(t)}$, which is a Floquet state for the driving with the instantaneous
 parameters ${\bf R}(t)$.
The Floquet state $\ket{\Psi_{\gamma,\,{\bf R}(t)}(t)}$ is defined as the adiabatic continuation of the initial Floquet state: 
we will explain below how to construct it in our case. 
In a close analogy to the standard adiabatic
theorem, the Floquet adiabatic evolution is valid if $t_f$ is much larger than the inverse gap in the Floquet spectrum.
So, if there is a time $t^*$ in the evolution and a quasi-energy $\overline{\mu}_\beta(t^*)$ such that $t_f\sim 1/(\overline{\mu}_\beta(t^*)-\overline{\mu}_\gamma(t^*))$, the dynamics for $t\gtrsim t^*$ deviates from the adiabatic continuation. It is important that, at all times,
$t_f$ is much larger than the instantaneous inverse frequency: in this way the driving can be considered locally periodic and the use of Floquet states is meaningful.}

{In our protocol, we start our evolution with a driving whose frequency $\Omega$ is very high. In this limit, the time-averaged Hamiltonian
coincides with the Floquet Hamiltonian~\cite{Russomanno_PRL12}, its GS is therefore a legitimate Floquet state~\footnote{Being constant in time is a particular case of periodic time-dependence.}. We take this state as the initial state of the dynamics: we slowly decrease the driving frequency and we follow its evolution. 
Thanks to the Floquet adiabatic theorem, the evolution will follow the adiabatic continuation of the initial state
until the gap in the Floquet spectrum becomes too small. We give the definition of Floquet ground state (FGS)
to this adiabatic continuation of the high-frequency ground state.
We emphasize that, in general, the FGS is not the ground state of any Hamiltonian\footnote{For instance, it is not the ground state of the Floquet Hamiltonian $\hat{H}_{F,\Omega}(t)$, because of the folding of the quasi-energies in the first Brillouin zone.}. We stress again that
the FGS is the adiabatic continuation at finite
frequency of the GS of the high-frequency average Hamiltonian; that is why it inherits the word ``ground''. As we will show, it inherits also many interesting properties.}

{Before doing that, being this adiabatic continuation so important, we will
give some notion concerning its formal construction. We fix the phase of the periodic motion $\theta\equiv t/\tau\in[0,2\pi]$,
 we take a Floquet mode~\footnote{The Floquet mode $\ket{\Phi_{\gamma}(t)}$ is equal to the Floquet state $\ket{\Psi_{\gamma}(t)}$ up to a phase, but it has the advantage of being time-periodic. If we know the Floquet modes we usually know also the quasi-energies (they are respectively eigenvectors and eigenvalues of the same operator, as we have explained above); if needed we can obtain the Floquet state
by inserting the phase factor $\nep^{-i\mu_{\gamma} t}$.
} 
at frequency $\Omega$ $\ket{\Phi_{\gamma,\,\Omega}(\tau\theta)}$ and we give 
the prescription for its adiabatic continuation for an infinitesimal change of frequency. If we take a frequency infinitesimally
smaller $\Omega-\delta\Omega$, the adiabatic continuation is the Floquet mode at the new frequency 
$\ket{\Phi_{\eta,\Omega-\delta\Omega}\left(\tau\left(1+\frac{\Delta\Omega}{\Omega}\right)\theta\right)}$ maximizing the overlap
$\left|\left\langle\Phi_{\eta,\Omega-\delta\Omega}\left(\tau\left(1+\frac{\Delta\Omega}{\Omega}\right)\theta\right)\right.\ket{\Phi_{\gamma,\Omega}(\tau\theta)}\right|^2$. ``Integrating'' this infinitesimal step from  $\Omega=\infty$ (where we initialize the state as the GS of the time-averaged Hamiltonian) to the desired value of $\Omega$, and repeating for all the values
of $\theta\in[0,2\pi]$ we obtain the full time-dependence of the Floquet ground mode $\ket{\Phi_{{\rm FGS}\,\Omega}(t)}$ (the Floquet ground state $\ket{\Psi_{{\rm FGS}\,\Omega}(t)}$ is obtained by
multiplying the phase factor $\nep^{-i\overline{\mu}_{\rm FGS} t}$, where $\overline{\mu}_{\rm FGS}$ is the corresponding quasi-energy). 
We are not able to do this adiabatic continuation in general, but only in the case of an integrable
system where we have access to all the Floquet modes and quasi-energies. In the integrable spin chain we explicitly discuss, we numerically iterate
the elementary step, starting from the GS of the time-averaged Hamiltonian
at a value of $\Omega$ very high compared to the spectral width of the elementary excitations. In this way we are going to show that, in analogy to the ground state of static systems, the FGS of a physical
system can undergo (non-equilibrium) quantum phase transitions.
}

We consider a homogeneous quantum Ising chain (see Ref.~\cite{simon2011quantum,kim2010quantum,Labrutt:preprint2015} for the experimental realization of this model with trapped atoms and ions): 
\begin{equation}  \label{hamil}
	\hat{H}^{(\alpha)}(t) = -\frac{1}{2} \sum_{j=1}^{L} \left[ J  \hat{\sigma}_j^z \hat{\sigma}_{j+1}^z + h(t) \hat{\sigma}_j^x 
             + \alpha \,\hat{\sigma}_j^z\hat{\sigma}_{j+2}^z\right]  \;.
\end{equation}
Here, the $\hat{\sigma}^{x,z}_j$ are spins (Pauli matrices) at site $j$ of a chain of length $L$, 
the $J$ is a longitudinal nearest neighbour coupling, the $\alpha$ a longitudinal next-nearest-neighbour coupling
and the $h(t)$ a transverse field. 
%
We assume in all the paper $J=1$ and start our analysis by considering $\alpha=0$. Although at this point the model is integrable, it 
describes many universal properties of 1d models with the
same symmetry: as we will see, {we propose that the same can happen for the properties of the FGS in the non-equilibrium case. }
Through a Jordan-Wigner transformation~\cite{Lieb_AP61}, $\hat{H}^{(0)}(t)$ can be reduced
to an integrable quadratic-fermion form. 
(see Refs.~\cite{Lieb_AP61,russomanno_JSTAT15,Russomanno_PRL12,Russomanno_JSTAT13,Calabrese_JSTAT12a,Calabrese_JSTAT12b,Batisdas_PRA12,Caneva_PRB07,Cherng_PRA06} for details).
After a Fourier transform, the Hamiltonian can be decomposed in $L/2$ $2\times 2$ blocks of the form
$\hat{H}_k^{(0)}(t)=\left[ \epsilon_k\left(h(t)\right)  \left(\opcdag{k} \opc{k}-\opc{-k} \opcdag{-k}\right) - i\Delta_k \left(\opcdag{k} \opcdag{-k}-\opc{-k} \opc{k}\right)\right]$ 
($\opc{k}$ are fermionic destruction operator) and 
%
%
%
the time-dependent Schr\"odinger equation is solved by a 
BCS-like state {$\ket{\Psi(t)} = \prod_{k>0}\left(v_k(t)+u_k(t)\opcdag{k} \opcdag{-k}\right)\ket{0}$}. 
\begin{figure}
%
\includegraphics[width=8.5cm]{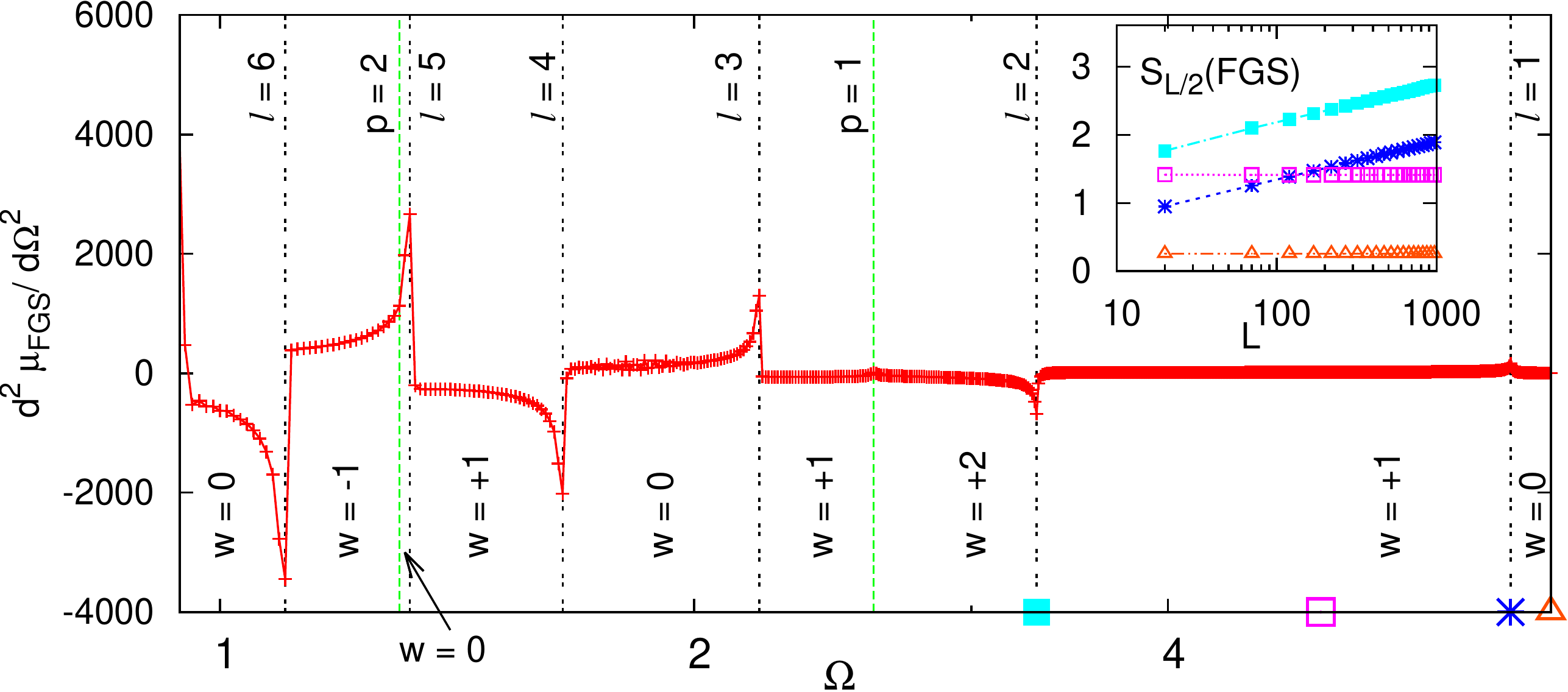}
\caption{
Second derivative of the Floquet ground state (FGS) quasi-energy versus the driving frequency, displaying second-order divergences in correspondence to each non-equilibrium transition. The labels $l$ and $p$ identify the transitions, and the winding-number $w$ sets the topological order parameter of the different phases. (Inset) Half-chain entanglement entropy of the FGS versus the system size $L$ (the considered frequencies are indicated by the
corresponding points in the horizontal axis of the main figure). $S_{L/2}({\rm FGS})$ grows logarithmically at the transitions, and follows an area law (saturates for large $L$) elsewhere. (Numerical parameters: $h_0=2.3$, $A=1.0$, $L=200$ in the main figure, periodic boundary conditions in the spins, antiperiodic in the fermions.)}

\label{noneq-trans:fig}
\end{figure}
%
%
We numerically calculate, for each $k$, the
two single-particle Floquet modes~\cite{Russomanno_PRL12,Russomanno_JSTAT13}
$\ket{\phi_k^\pm(t)}=\left(v_k^\pm(t)+u_k^\pm(t)\opcdag{k} \opcdag{-k}\right)\ket{0}$ {(with $u_k^+(t)=(v_k^-(t))^*$ and $v_k^+(t)=-u_k^-(t))^*$ $\tau$-periodic)} and the quasi-energies (in the first Brillouin zone)
$\pm\mu_k$. {It is important to notice that we can define a time-periodic creation operator $\opgammadag{P\,k}(t)=u_k^+(t)\opcdag{k}+v_k^+(t)\opc{-k}$
such that $\ket{\phi_k^+(t)}=\opgammadag{P\,k}(t)\opgammadag{P\,-k}(t)\ket{\phi_k^-(t)}$: one Floquet mode can be obtained from the other by
adding two quasi-particle operators.} 
{Numerically applying the adiabatic continuation prescription, we find that,} for this model, we can
explicitly evaluate the FGS: {its corresponding Floquet mode} has the BCS-like form 
$\ket{\Phi_{\rm FGS}(t)}=\prod_{k>0}\ket{\phi_k^-(t)}$ with quasi-energy 
$\overline{\mu}_{\rm FGS}=-\sum_{k>0}\mu_k$. {In this lucky case, $\ket{\Phi_{\rm FGS}(t)}$ coincides 
with the ground state of the instantaneous Floquet Hamiltonian taken in the first Brillouin zone.} 

This model shows a series of interesting Floquet resonances~\cite{russomanno_JSTAT15,Batisdas_PRA12}. 
They are discussed in detail in Ref.~\cite{russomanno_JSTAT15}, where it is shown that their positions 
depend only on the
time-averaged value $h_0$ of the periodic drive $h(t)$. 
They are associated to many-photon resonances  and occur only at $k=0,~\pi$.
The $k=0$-resonances occur when $2\left|h_0-1\right|=p\,\Omega$, the $k=\pi$-ones if $2\left|h_0+1\right|=l\Omega$ 
($p,~l\in\mathbb{Z}$ is the order of the resonance -- see Refs.~\cite{russomanno_JSTAT15,russomanno_16} for more details).
\begin{figure}
%
%
   \begin{tabular}{c}
   \hspace{0.cm}\resizebox{80mm}{!}{\includegraphics{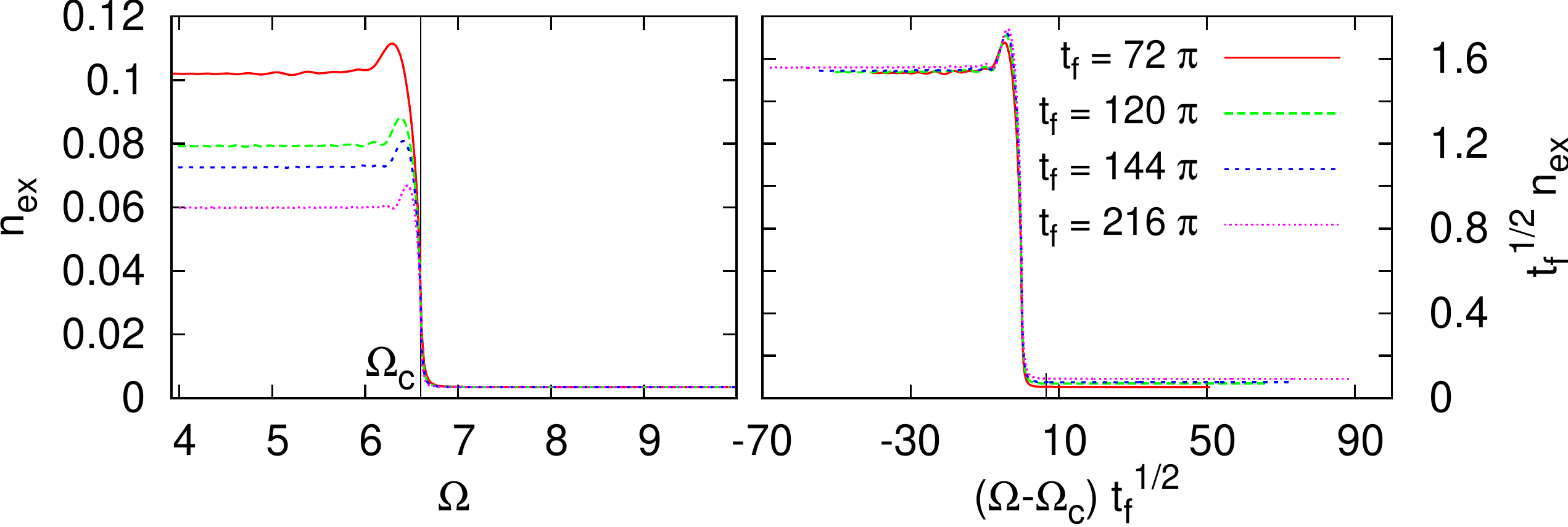}}\\
%
%
   \end{tabular}
%
\caption{
(Left panel) Number of excitations vs $\Omega$ for different values of $t_f$: it starts increasing after crossing the
transition. (Right panel) Kibble-Zurek scaling of the number
of excitations: the rescaled curves for different $t_f$ overlap after the rescaling (the overlap is not perfect because we initialize
the system in the ground state of the time-averaged Hamiltonian, see SM of Ref.~\cite{KZ_arx} for details). 
(Numerical parameters: $h_0=2.3,~A=1$,~$L=300,~\Omega_i=10,~\Omega_f=4$; periodic boundary conditions in the spins, antiperiodic
in the fermions.)} 
%
\label{transismo_ent:fig}
\end{figure}

In this work we find that the resonances
actually mark a series of non-equilibrium quantum phase transitions
where the properties of the FGS change. To see that, we compute $S_{L/2}({\rm FGS})$, the half-chain entanglement entropy of $\ket{\Phi_{\rm FGS}(0)}$. 
We find that $S_{L/2}({\rm FGS})$ grows logarithmically with $L$ at the resonance points while it remains constant elsewhere (see the inset of Fig.~\ref{noneq-trans:fig}); the same happens 
at equilibrium for the {ground-state quantum phase transition}~\cite{Latorre_QIC04,amico08}.  Remarkably the quasi-energy is {\em non-analytic} at the resonance points: 
its second derivative with respect to $\Omega$ diverges (see Fig.~\ref{noneq-trans:fig}). This means that at these points the Floquet Hamiltonian undergoes
 second order quantum phase transitions~\cite{Sachdev:book}. 
In order to simplify the numerical analysis, we obtain the results in Fig.~\ref{noneq-trans:fig} applying a square-wave driving of the form
$h(t)=h_0+A\sign\left(\sin(\Omega t)\right)$ 
However, we stress that the specific form of the driving does not affect
the position in $\Omega$ of the resonances -- which only depends on $h_0$ -- and the nature of the
non-analiticities occurring there.
{In a future publication, we will discuss these transitions in the spin representation, showing that the different phases 
are characterized by a local or string order parameter~\cite{figata}.
Here, we focus on the fermionic representation, where the transitions are 
of topological nature. We consider the time-reversal invariant points $t_a=\pi/(2\Omega)$ and $t_b=3\pi/(2\Omega)$~\cite{Erez_PRB10}, where
the Floquet Hamiltonian falls in the $\mathbb{Z}$ symmetry class (see~\cite{Kane_RMP}). The different phases are characterized by 
a $\mathbb{Z}$} topological invariant, the winding number $w$ (see Ref.~\cite{topological_notes} and Fig.~\ref{noneq-trans:fig}). 
The time-reversal invariant Floquet Hamiltonian $\hat{H}_{F,\Omega}^{(0)}(t_a=\pi/(2\Omega))$ can be factorized in $2\times 2$ blocks which can be represented as $f_y(k)\sigma_y+f_z(k)\sigma_z$ (thanks to time-reversal invariance, there is no
component along $\sigma_x$); defining $f(k)\equiv f_y(k)+if_z(k)$, the winding number is defined as~\cite{topological_notes}
%
  $w\equiv\frac{1}{2\pi i}\int_{-\pi}^{\pi}\ud k \frac{d}{dk}\log f(k)$. From a geometric perspective, $w$ is the number of times the vector $\left(\begin{array}{cc}f_y(k),&f_z(k) \end{array}\right)$
turns around the origin $\left(\begin{array}{cc}0,&0 \end{array}\right)$ in counterclockwise sense, when $k$ moves from $-\pi$ to $\pi$.
%
{Interestingly, if we consider the FGS at $t\neq t_a,t_b$, time-reversal symmetry is broken, the Floquet Hamiltonian falls in the $\mathbb{Z}_2$
symmetry class~\cite{Kane_RMP}: the transitions are in the same points of the parameter space, but the phases are
characterized by a different topological invariant, $\nu$ defined in~\cite{Alicea_RPP12}, which, 
in the sequence of different phases, is alternately 1 and -1. Therefore, the symmetries of the FGS and the 
topological invariant characterizing it are different
according to the considered time $t$ of the periodic driving.}
These transitions are reminiscent of those occurring in Floquet 
topological insulators~\cite{Lindner_Nat11,Jotzu_Nat14,Oca_PRB79,LS_15} 
and are a manifestation of a genuine many-body phenomenon: we will see that 
they persist in presence of interactions.

%
The linear crossing of one of these non-equilibrium transitions leads to {the breaking of the adiabatic evolution (the gap in the Floquet
spectrum closes at these points): this gives rise to} a K-Z-scaling phenomenon which we define ``Floquet Kibble-Zurek''. 
In order to observe it, we apply to the system 
a time-dependent drive of the form $h(t) = h_0 + A \sign\left[\sin( \phi(t) ) \right]$ where we linearly ramp the 
instantaneous frequency, $\Omega(t) = \frac{d}{dt}\phi(t)=\Omega_i + (\Omega_f-\Omega_i)\frac{t}{t_f}$. 
We specifically choose $\Omega_f$ 
below $\Omega_c$, but above the other transitions and $\Omega_i$ quite larger than $\Omega_c$.
$\Omega_i$ is large enough that the ground state of the time-averaged Hamiltonian 
is a very good approximation for the FGS:
we prepare the system in this ground state and we ramp it in a time $t_f$ to $\Omega_f$:
We cross the first ($l=1$) non-equilibrium transition at frequency $\Omega_c$.
The adiabatic limit of this protocol is attained when $t_f\to\infty$, implying that $t_f$ is our adiabaticity parameter.
{Being $t_f\gg2\pi/\Omega_f,2\pi/\Omega_i$, we can approximate the driving as locally periodic, with a period which slowly
changes from one cycle to the next, and apply Floquet adiabatic theorem. To make
our analysis simpler, in our numerics we will focus on the times $t_n$ where the instantaneous phase of the locally periodic driving is 0: these times are given
by $\phi(t_n)=2\pi n$.}

\begin{figure}
\centering
\begin{tabular}{c}
\resizebox{80mm}{!}{\includegraphics[width=8.5cm]{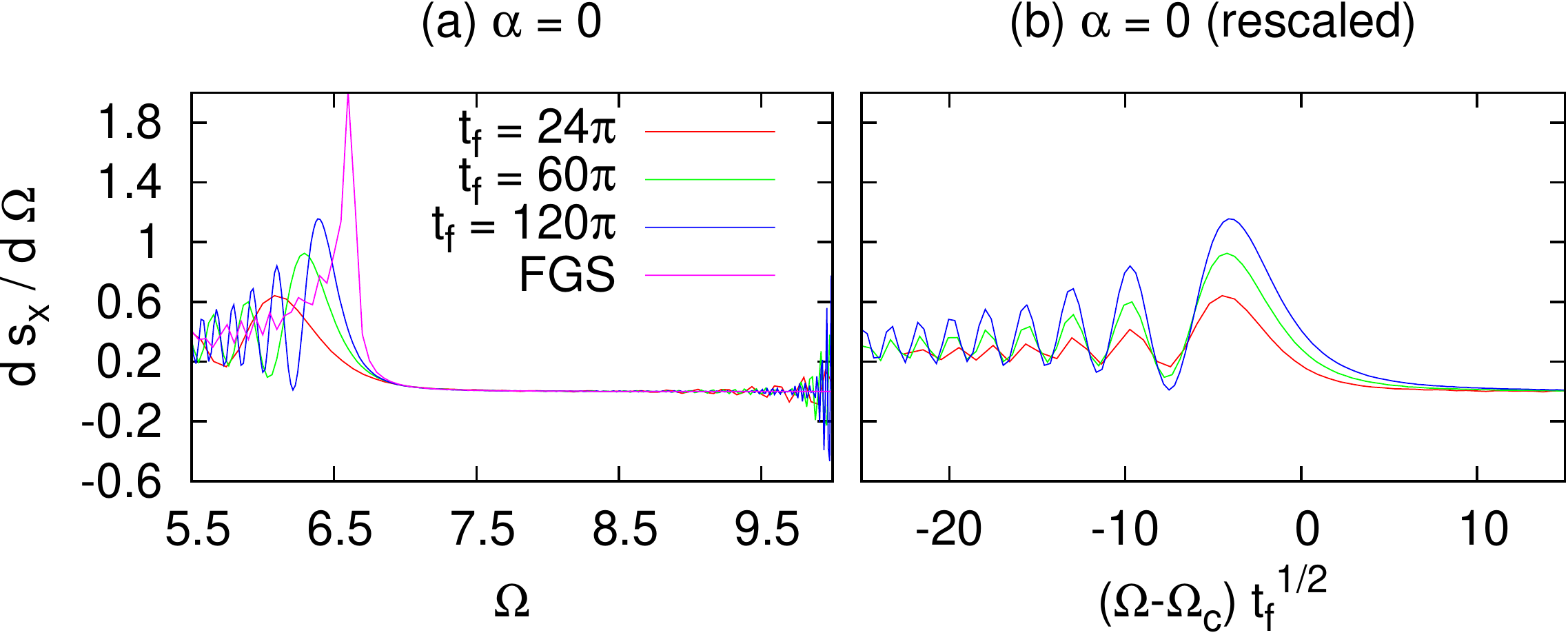}}\\
\resizebox{80mm}{!}{\includegraphics[width=8cm]{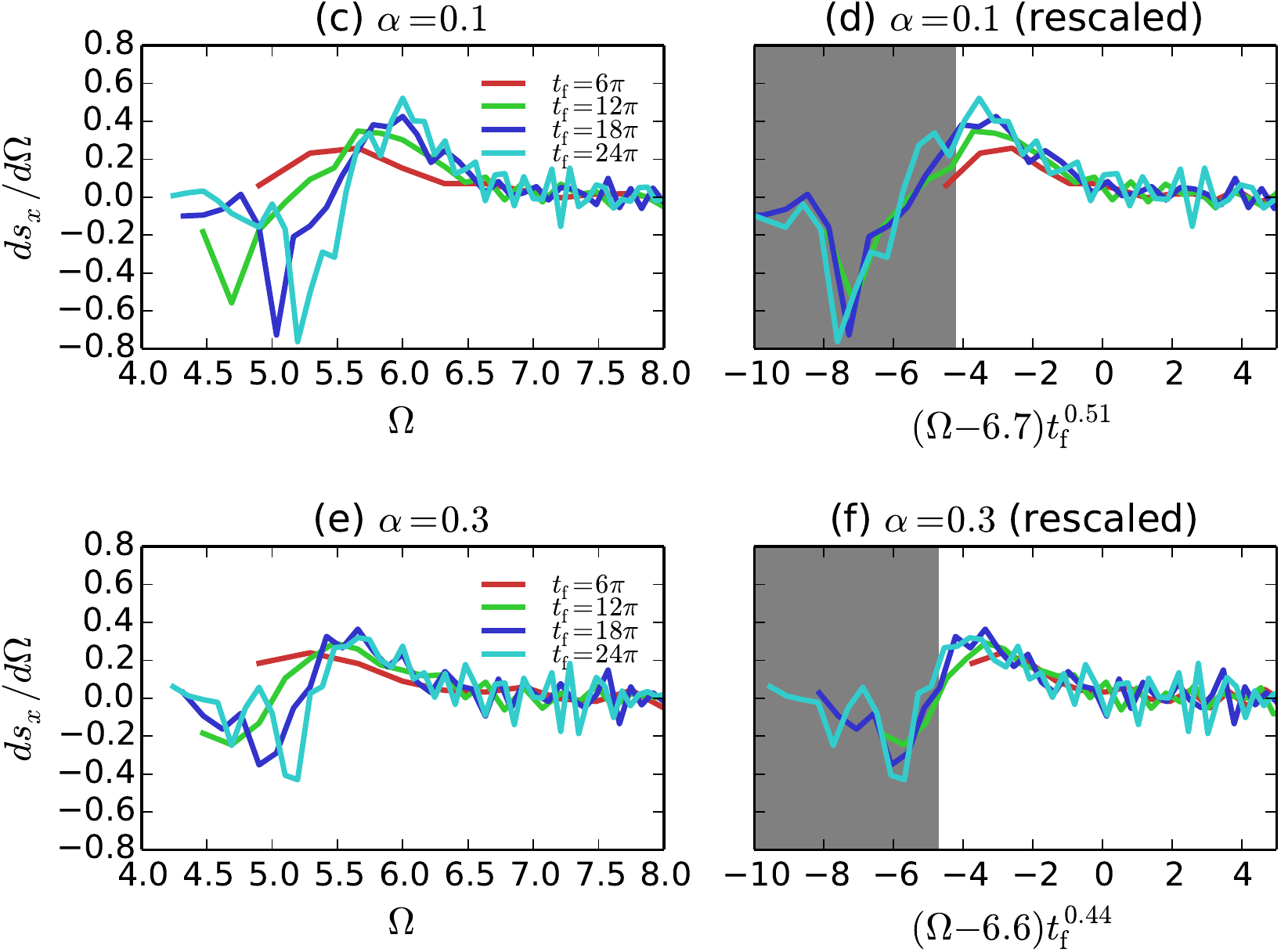}}
\end{tabular}
\caption{K-Z scaling of the transverse magnetization derivative crossing a non-equilibrium quantum phase transition. (a-b) 
At the integrable point $\alpha=0$ the system allows for an exact solution and displays a clear scaling behavior close to the resonance (at $\Omega_c=6.6$); we plot also the FGS value diverging at the transition. 
(c-d) For small values of the integrability breaking term $\alpha=0.1$, an exact solution is not available, but our numerical results are consistent with a K-Z scaling. (e-f) Far from integrability, at $\alpha=0.3$, the K-Z scaling still persists with approximately the same exponent. The values
of $\Omega_c(\alpha)$ and $\beta(\alpha)$ used in subplots (b) and (d) were found by maximizing the distance between the rescaled functions. In the gray regions the truncation error is large (see SM of Ref.~\cite{KZ_arx}) (Numerical parameters: $h_0=2.3,~A=1,~M=30,~\Omega_i=10,~\Omega_f=4$, open boundary conditions. $L=300$ in panels (a-b), $L=100$ in
panels (c-f).) 
} 
\label{transismo_alpha:fig}
\end{figure}

We can see that, beyond the critical point $\Omega_c$ -- {where the gap in the Floquet spectrum closes}, adiabaticity breaks down and 
fluctuations
cannot adiabatically follow the FGS; 
this results in the creation of excitations whose density {is defined as 
$n_{\rm ex}(t)=\frac{1}{L}\bra{\Psi(t)}\sum_{k>0}\opgammadag{P\,k}(t)\opgamma{P\,k}(t)\ket{\Psi(t)}$. The sum extends
over the $L/2$ values of $k$ allowed by the chosen periodic boundary conditions in the spins (antiperiodic in the fermions)~\cite{Russomanno_PRL12}; the creation operators and the state $\ket{\Psi(t)}$ solution of the Schr\"odinger equation have been defined in the discussion
above\footnote{As we will present in~\cite{figata}, the phase below $\Omega_c$ is characterized in the spin representation by a local
order parameter: the staggered magnetization along the $z$ axis. Therefore, excitations have the physical meaning of defects as
in the standard Kibble-Zurek phenomenon~\cite{Polkovnikov_RMP11,Zurek_PRL05}.}.} 
We show numerical results for $n_{\rm ex}$ in Fig.~\ref{transismo_ent:fig}: we see a very clear K-Z scaling of the form
$ n_{\rm ex} = t_f^{-1/2} \mathcal{N}\left((\Omega_c-\Omega)t_f^{1/2}\right)$. 
%
%
%
A similar scaling is obeyed by the energy difference with respect to the FGS. To observe these quantities it is however necessary to compare
the dynamics with the FGS, which is generically not accessible in experiments and numerical methods like t-DMRG.
{In contrast, objects of experimental interest show at most a scaling along the $\Omega-\Omega_c$ axis. 
For instance, }
the derivative of the (doubled) transverse magnetization $d s_x/d\Omega$ ($\hat{s}_x = 2\hat{m}_x=\frac{1}{L}\sum_j\hat{\sigma}_j^x$)
shows very clear oscillations which obey a ``partial'' scaling law in $t_f$. Indeed, the nodes of the oscillations
overlap if the results for different $t_f$ are plotted vs $\left(\Omega-\Omega_{c}\right)t_f^{1/2}$,
but the {oscillation amplitudes} do not show a clear scaling (see Fig.~\ref{transismo_alpha:fig}(a),(b)). 

%
To study the non-integrable case ($\alpha\neq 0$ in Eq.~\eqref{hamil}) we employ the openMPS package~\cite{endnote,Carr_NJP06} to directly compute the time evolution of the system. As in the integrable case, we initialize the system in the ground state of the time-averaged Hamiltonian. 
In contrast to the case of sudden quenches where the entanglement entropy and the truncation error rapidly grow in time, we find that for our protocol both quantities remain very small for long times (see SM of Ref.~\cite{KZ_arx} for details). To observe the Floquet Kibble-Zurek scaling, we 
still plot 
$ds_x/d\Omega$ as a function of the driving frequency $\Omega$, for different values of the adiabatic parameter $t_f$. As shown in Fig.~\ref{transismo_alpha:fig}, {each curve oscillates differently} but the nodes of
all the curves overlap if plotted versus $(\Omega-\Omega_c) t_f^\beta$. 

The two fitting parameters $\Omega_c$ and $t_f$ are chosen to minimize the distance between the rescaled curves (see SM of Ref.~\cite{KZ_arx} for details). Interestingly, the overlap between the curves extends beyond the strict validity of the numerical precision, in the region of large truncation error (gray areas in Fig.~\ref{transismo_alpha:fig}), and provides further indications of the correctness of the scaling. 
By comparing different fitting procedures we obtain that for $\alpha=0.1$, $\Omega_c=6.78\pm0.16$ and $\beta=0.51\pm0.09$, while for $\alpha=0.3$, $\Omega_c=6.6\pm0.3$ and $\beta=0.44\pm0.09$. These values are consistent with a universal scaling with the same exponent as in the integrable case, $\beta=0.5$, but the large error-bars do not allow us to make a strong statement. Note that 
in one dimension there are cases of topological transitions at equilibrium with continuously varying critical exponents (see for example Ref.~\cite{Schultz_PRB86,Kazzuo_93,Berg_08}). To address this question it is necessary to improve the quality of the numerical method (we keep here only up to $M=30$ eigenstates of the density matrix) 
by employing parallelized algorithms, which go beyond the scope of the present paper.

%
In conclusion, we find -- in a many-body system -- a low-entanglement Floquet state analogous to the ground state.
We adiabatically follow it until a many-body Floquet resonance occurs: the low entanglement gives the possibility to
numerically study non-integrable models by means of t-DMRG.
We explicitly discuss the case of an integrable model where the many-body resonances are non-equilibrium phase transitions of the FGS. 
We see a strict analog of the K-Z scaling when crossing the highest of these transitions: this phenomenon allows us to identify 
a similar effect for moderate non-integrabilities as well. This is highly non-trivial, considering the high degree of folding and
mixing of the Floquet spectrum of a driven interacting many-body system.
We find that the integrable transitions have topological properties which should persist for weak interactions. The persistence of the
critical behaviour for moderate interactions suggests the absence 
of many-body delocalization of the FGS. Although other parts of the Floquet
spectrum may undergo many-body delocalization~\cite{Dalessio_AP13,Rigol_PRX14,Ponte_PRL15,Ponte_AP15,Emanuele_2014:preprint,Rosch_15}, our method does not give the possibility to directly observe it.
Our results were obtained by means of limited resources~\cite{endnote}: the resulting uncertainty
does not allow us to clarify what happens in our system for large interactions. 
Further work is therefore needed to 
see if the FGS itself can undergo many-body delocalization in our model and in other interesting 1d 
systems~\cite{Rigol_PRX14,Rosch_15,Emanuele_2014:preprint,Dalessio_AP13,Ponte_PRL15,Ponte_AP15}.

\acknowledgements

We acknowledge useful discussions with E. Berg, R. Berkovitz, E. Demler and R. Fazio. Research was supported by the Coleman-Soref foundation,
by the Israeli Science Foundation (grant number $1542/14$) and by the EU FP7 under grant agreement n. 641122 (STREP-QUIC). This work was not supported by any military agency.


\end{document}